\documentclass[prl,a4paper,twocolumn,superscriptaddress,preprintnumbers]{revtex4-1}

\usepackage[utf8]{inputenc}
\usepackage{amssymb}
\usepackage{amsmath}
\usepackage{graphicx}
\usepackage{psfrag}
\usepackage{latexsym}
\usepackage{hyperref}

\newcommand{\bea}{\begin{eqnarray}}
\newcommand{\beal}[1]{\begin{eqnarray}\label{#1}}
\newcommand{\eea}{\end{eqnarray}}
\newcommand{\be}{\begin{equation}}
\newcommand{\bel}[1]{\begin{equation}\label{#1}}
\newcommand{\ee}{\end{equation}}
\newcommand{\rf}[1]{Eq.~(\ref{#1})}
\newcommand{\rff}[1]{Fig.~(\ref{#1})}
\newcommand{\rfc}[1]{Ref.~\cite{#1}}
\newcommand{\f}[2]{\frac{#1}{#2}}

\newcommand{\bit}{\begin{itemize}}
\newcommand{\eit}{\end{itemize}}
\newcommand{\ben}{\begin{enumerate}}
\newcommand{\een}{\end{enumerate}}

\newcommand{\symm}{${\mathcal N}=4$ SYM}

\newcommand{\pT}{{\cal P}_T}
\newcommand{\pL}{{\cal P}_L}

\newcommand{\ed}{{\cal E}}
\newcommand{\reltime}{{\tau_{rel}}}

\newcommand{\pa}{{\cal A}}
\newcommand{\pac}{{a}}

\begin{document}

\title{Hydrodynamization in kinetic theory:\\
Transient modes and the gradient expansion}
\preprint{CERN-TH-2016-199}

\author{Michal P. Heller}
\email{michal.p.heller@aei.mpg.de}
\affiliation{Max Planck Institute for Gravitational Physics, Potsdam-Golm, D-14476, Germany}
\affiliation{Perimeter Institute for Theoretical Physics, Waterloo, Ontario N2L 2Y5, Canada}
\affiliation{National Centre for Nuclear Research,
  00-681 Warsaw, Poland}

\author{Aleksi Kurkela}
\email{aleksi.kurkela@cern.ch}
\affiliation{Theoretical Physics Department, CERN, Geneva, Switzerland}
\affiliation{Faculty of Science and Technology, University of Stavanger, 4036 Stavanger, Norway}

\author{Micha\l\ Spali\'nski}
\email{michal.spalinski@ncbj.gov.pl}
\affiliation{National Centre for Nuclear Research,
  00-681 Warsaw, Poland}
\affiliation{Physics Department, University of Bia{\l}ystok,
  15-245 Bia\l ystok, Poland}

\author{Viktor Svensson}
\email{viktor.svensson@aei.mpg.de}
\affiliation{National Centre for Nuclear Research,
  00-681 Warsaw, Poland}
\affiliation{Max Planck Institute for Gravitational Physics, Potsdam-Golm, D-14476, Germany}

\begin{abstract}

We explore the transition to hydrodynamics in a weakly-coupled model of quark-gluon plasma given by kinetic theory in the relaxation time approximation with conformal symmetry. We demonstrate that the gradient expansion in this model has a vanishing radius of convergence due to the presence of a transient (nonhydrodynamic) mode, in a way similar to results obtained earlier in strongly-coupled gauge theories. This suggests that the mechanism by which hydrodynamic behaviour emerges is the same, which we further corroborate by a novel comparison between solutions of different weakly and strongly coupled models. However, in contrast with other known cases, we find that not all the singularities of the analytic continuation of the Borel transform of the gradient expansion correspond to transient excitations of the microscopic system: some of them reflect analytic properties of the kinetic equation when the proper time is continued to complex values.

\end{abstract}

\maketitle

\noindent \emph{1. Introduction and summary.--} Heavy-ion collisions at RHIC and
LHC provide an outstanding opportunity to test our understanding of QCD.
Perhaps unsurprisingly, a fully ab initio theoretical
description has turned out to be very challenging. This has led to the
exploration of models of increasing complexity and often non-overlapping
domains of validity -- see, e.g.,
Refs.~\cite{Florkowski:2017olj,Romatschke:2017ejr} for a review of such models
in the context of the hydrodynamic description in ultrarelativistic heavy-ion
collisions.

In this Letter we focus on the poorly understood transient far-from-equilibrium regime, which precedes viscous
hydrodynamic evolution of quark-gluon plasma (QGP). There are two approaches to the study of the transition to
hydrodynamics (hydrodynamization)  in non-Abelian gauge theories like QCD: a weakly-coupled description based on
effective kinetic theory (EKT) \cite{Arnold:2002zm} (see also
Refs.~\cite{Baier:2000sb,Arnold:2006fz,Kurkela:2011ti,Kurkela:2011ub,Kurkela:2014tea,Kurkela:2015qoa}) and a
strongly-coupled plasma paradigm based on holography~\cite{Maldacena:1997re} (see
Refs.~\cite{Chesler:2009cy,Chesler:2010bi,Heller:2011ju,CasalderreySolana:2011us} for sample results).
They involve
very different physical pictures and mathematical frameworks: the first relies on the Boltzmann equation while the
second makes use of higher-dimensional Einstein equations. Since under experimental conditions the QCD coupling is
neither parametrically small nor large, it is crucial to understand which implications of these approaches can be viewed
as universal.

Our aim is to shed light on equilibration in weakly-coupled systems by
examining large-order behaviour of the hydrodynamic gradient
expansion~\cite{Heller:2013fn} in the framework of the
kinetic theory model given by the Boltzmann equation in the relaxation time
approximation (RTA)~\cite{Anderson}.
We also assume conformal symmetry, as its
breaking at strong coupling does not significantly alter equilibration
processes~\cite{Buchel:2015saa,Janik:2015waa}, whereas at weak coupling at
vanishing quark masses it is a NLO effect (see,
e.g.,~Ref.~\cite{Arnold:2006fz}). The key result of this Letter is demonstrating the vanishing radius of convergence of the gradient expansion in this
kinetic theory model (see also Ref.~\cite{Denicol:2016bjh} which studied in this context the RTA kinetic theory with constant relaxation time) and understanding some puzzling features revealed by these studies. The reason for the divergence turns
out to be the same as in the case of holographic plasma: the presence of fast-decaying (nonhydrodynamic)
modes~\cite{Baier:2007ix,Heller:2015dha,Romatschke:2015gic,Kurkela:2017xis} whose relaxation controls the emergence of
hydrodynamic behaviour (and in particular, its applicability to the physics of heavy-ion
collisions~\cite{Romatschke:2015gxa,Habich:2015rtj,Spalinski:2016fnj}).
This is connected with the existence of attractors which govern the
evolution far from equilibrium~\cite{Heller:2015dha,Romatschke:2016hle,Romatschke:2017vte,Spalinski:2017mel,Romatschke:2017acs,Strickland:2017kux,Florkowski:2017jnz,Denicol:2017lxn,Casalderrey-Solana:2017zyh,Blaizot:2017ucy}.
In this context, see also
Refs.~\cite{Brewer:2015ipa,Bantilan:2016qos} for studies of possible manifestations of analogous fast-decaying modes in
trapped Fermi gases close to unitarity, as well as Ref.~\cite{Grozdanov:2016vgg} where various features of transient
modes are analyzed as a function of the microscopic interaction strength
in a holographic toy model.

Quite remarkably, our analysis of the Borel transform of the gradient expansion in the RTA kinetic theory reveals not
only the expected purely decaying mode~\cite{Romatschke:2015gic}, but also singularities which could naively be interpreted as transient contributions to the
energy-momentum tensor exhibiting damped oscillatory behaviour, similar to the findings of Ref.~\cite{Bazow:2015zca}. This would however be surprising, since the mechanism described there does not apply to the RTA theory~\footnote{M.~Martinez and U.~Heinz, private communication.}. In fact, we demonstrate below that
these singularities are instead a manifestation of analytic properties of the evolution equations in
complexified time.
This feature is related to what has been observed in other contexts where large-order behavior of perturbative
series expansions is used to draw conclusions about non-perturbative effects
(see,
e.g., Refs.~\cite{Dunne,Dunne:2015eaa,Basar:2013eka,Aniceto:2018bis}). 

\begin{figure}[ht]
\center
\includegraphics[width=0.49\textwidth]{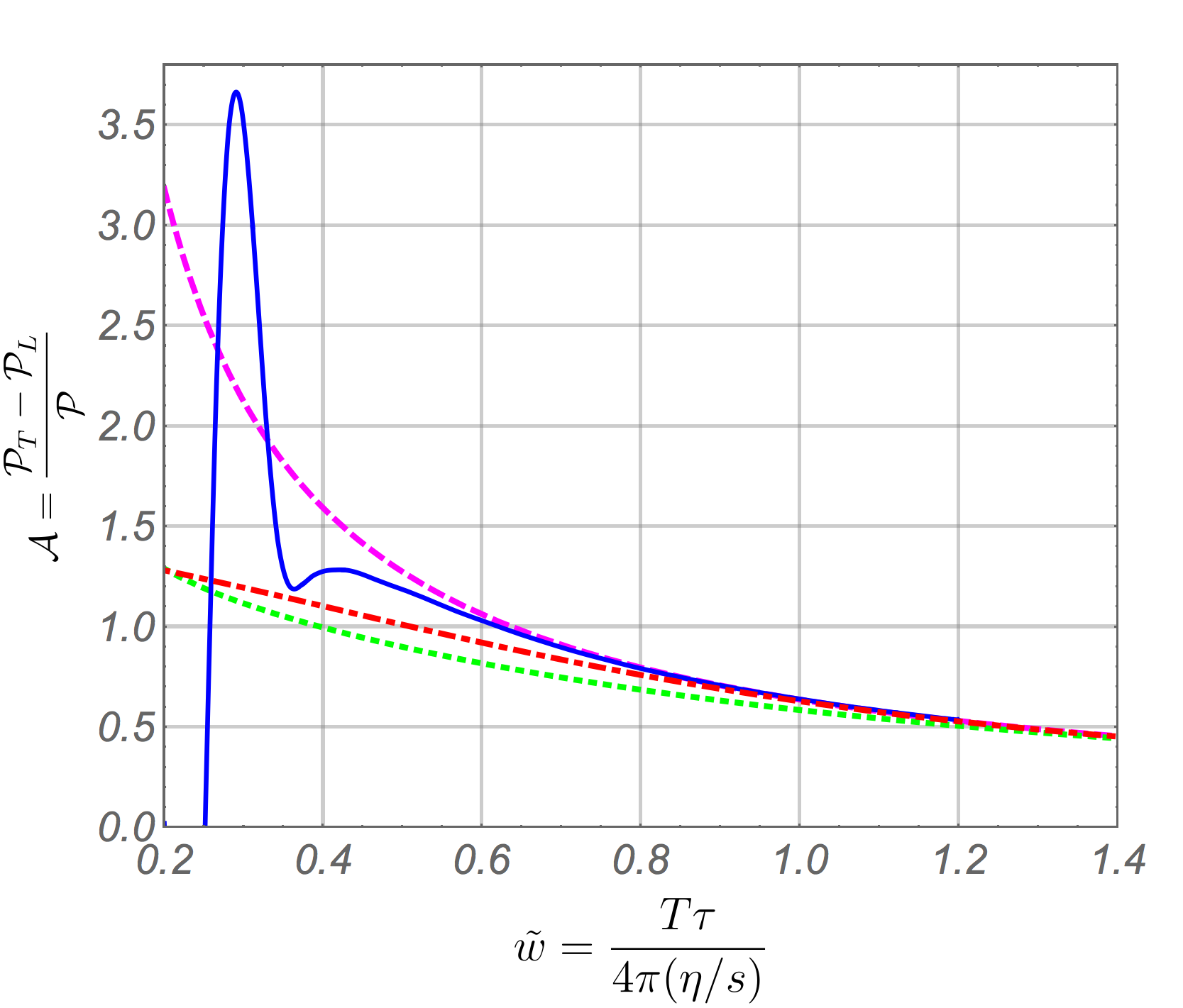}
\caption{ The dashed magenta curve represents first order hydrodynamics, the blue line is the holographic
  result, and the red, dashed-dotted line is from EKT. The green dotted curve stands for a solution of RTA
  starting
  from similar initial distribution as EKT and with the same shear viscosity, $\eta/s=0.624$. Despite
  differences in microscopic dynamics, one sees significant qualitative and some quantitative similarities
  between different theories.
  }
\label{fig:tilde}
\end{figure}

Our conclusions concerning hydrodynamization provide strong motivation for comparing numerical solutions of the RTA evolution
equations with the EKT results at intermediate coupling reported in Refs.~\cite{Kurkela:2014tea,Kurkela:2015qoa} and
with the AdS/CFT-based simulations of Ref.~\cite{Jankowski:2014lna}. We uncover semi-quantitative agreement and, as a
byproduct of this analysis, we present a new and effective way of visualizing (see Fig.~\ref{fig:tilde}) the correlation
between the hydrodynamization time and the value of the $\eta/s$ ratio noted in Ref.~\cite{Keegan:2015avk}.

\vspace{10 pt}

\noindent \emph{2. Kinetic theory.--} We address the issues discussed above in the
context of Bjorken flow \cite{Bjorken:1982qr}, which is conveniently
formulated in (proper time)-rapidity coordinates $\tau$-$y$. They are related
to Minkowski lab-frame coordinates $t$-$z$ by $t = \tau \cosh y$ and $z = \tau
\sinh y$ where $z$ is the collision axis. Assuming
translation symmetry in the transverse plane
$\mathbf{x}_{T}$, the on-shell distribution function $f$
depends only on the proper time $\tau$, the modulus of the transverse momentum
$p\equiv |\mathbf{p}_T|$ and the boost-invariant variable $u=\tau^2 p^y$.

In the RTA the collision kernel appearing in the Boltzmann equation is linearized
around the the equilibrium distribution which for simplicity we take to be Boltzmann
\bel{boldist}
f_0(\tau,u,p) =
\frac{1}{(2\pi)^3} \exp\left[
- \frac{\sqrt{u^2+p^2 \tau^2}}{\tau \, T(\tau)}  \right].
\ee
The RTA Boltzmann equation takes the form
\bel{bebif}
\frac{\partial f(\tau, u, p)}{\partial \tau} = \frac{1}{\reltime} \left\{ f_0(\tau, u, p)
  -f(\tau, u, p) \right\}.
\ee
To ensure conformal symmetry, we assume that the relaxation time is of the form
\be
\label{eq.reltime}
\reltime=\frac{\gamma}{T(\tau)},
\ee
where $\gamma$ is dimensionless and is the only parameter of this
model. The dependence of temperature on the proper time is determined
dynamically by imposing the Landau matching condition~\cite{Baym:1984np,Florkowski:2013lya}
\bel{lmc}
\ed(\tau) = \frac{3}{\pi^{2}} T^4(\tau),
\ee
where
\be
\ed(\tau) =2  \int d^4p \, \delta \left( p^2\right) \theta (p^0)
\, \frac{u^2+p^{2}\tau^{2}}{\tau^2}\,  f(\tau,u,p)
\ee
is the energy density (per particle species).

\vspace{10 pt}
\noindent \emph{3. Hydrodynamic Gradient Expansion.--} In a conformal theory the
eigenvalues of the expectation value of the energy
momentum tensor in a boost-invariant state are
functions of the proper time $\tau$ alone. They are given by
the energy density $\ed$ and
the longitudinal and transverse pressures
$\pL$ and $\pT$:
\bel{PLT}
\pL = - \ed - \tau\, \dot{\ed}\ , \quad
\pT =  \ed + \f{1}{2} \tau\, \dot{\ed}
\ee
Away from equilibrium
$\pL$ and $\pT$ differ from the equilibrium pressure at the same energy density ${\cal P}\equiv{\cal E}/3$. It is convenient
to study the approach to equilibrium
by examining the behaviour of the pressure anisotropy
\bel{rdef}
\pa \equiv \f{\pT-\pL}{{\cal P}}
\ee
as a function of the dimensionless variable $w\equiv T\,\tau$.
The gradient expansion of $\pa$ takes the form
\bel{gradex}
\pa(w) = \sum_{n=1}^{\infty} \pac_n w^{-n} .
\ee
This follows directly
  from \rf{rdef} and \rf{PLT} if we use the fact that in
conformal theories near equilibrium $\ed \sim T^4$ and for
boost invariant
flow $T\sim\tau^{-1/3} +O(\tau^{-1})$ \cite{Bjorken:1982qr} (up to
exponentially suppressed corrections).

To determine the
coefficients $\pac_n$ we
look for a solution of the Boltzmann
equation \rf{bebif} in the form
\bel{fex}
f(\tau, u, p) = f_0(\tau, u, p) \left(1 +
\sum_{n=1}^{\infty}\,  w^{-n} \, h_n\left(\f{u}{w}, \f{p}{T}\right) \right) .
\ee
Inserting \rf{fex} into the Boltzmann equation \rf{bebif} one can algebraically determine the
functions $h_n$ in terms of the unknown coefficients $\pac_n$. A key step in
doing this is to eliminate proper time derivatives of temperature in favor of
$\pa$ and then using \rf{gradex}.

The Landau matching condition \rf{lmc} implies that at each order $n>0$
\bel{hnlmc}
\int d^4p \, \delta \left( p^2\right) \theta (p^0) \, \f{v^2}{\tau^2}\,
f_0(\tau,u,p) \, h_n\left( \f{u}{w},\f{p}{T}\right)
= 0 .
\ee
This condition amounts to a linear, algebraic equation which
determines the expansion coefficient $\pac_{n-1}$.

Proceeding this way we have calculated the expansion coefficients analytically
up to order 426 (we include the result of this calculation in the
supplemental material). As a cross-check, we have also calculated the first 10
terms using two independent, albeit slower, methods from Refs.~\cite{Jaiswal:2013npa} and~\cite{Baym:1984np,Florkowski:2013lya}, noting perfect agreement.

The leading expansion coefficients read
\be
\label{eq.r123}
\hspace{-5 pt}\pac_1=8/5\ \gamma, \quad \pac_2=32/105\ \gamma^2, \quad \pac_3 = - 416/525 \, \gamma^3 . \,
\ee
They can be used to match the transport coefficients of BRSSS
hydrodynamics~\cite{Baier:2007ix} to the RTA model. In particular, one finds
\be
\eta/s = \gamma/5.
\ee

At large orders, the expansion coefficients in \rf{gradex} exhibit factorial growth.  This demonstrates the
vanishing radius of convergence of the hydrodynamic series, which parallels similar findings obtained
numerically in \symm\ using holography \cite{Heller:2013fn,Buchel:2016cbj} as well as in hydrodynamics~\cite{Heller:2015dha,Basar:2015ava,Aniceto:2015mto}. This is to be expected on general
grounds, since the RTA theory contains, apart from hydrodynamic excitations, also a short-lived
mode~\cite{Romatschke:2015gic} whose physics is not captured by the truncated gradient expansion, as shown in
Refs.~\cite{Heller:2013fn,Heller:2015dha}.

\vspace{10 pt}

\noindent \emph{4. The Borel transform and short-lived modes.--} In this section we explore the Borel transform technique as a way to map out the excitations of expanding QGP. 

\begin{figure}[ht]
\center
\includegraphics[width=0.47\textwidth]{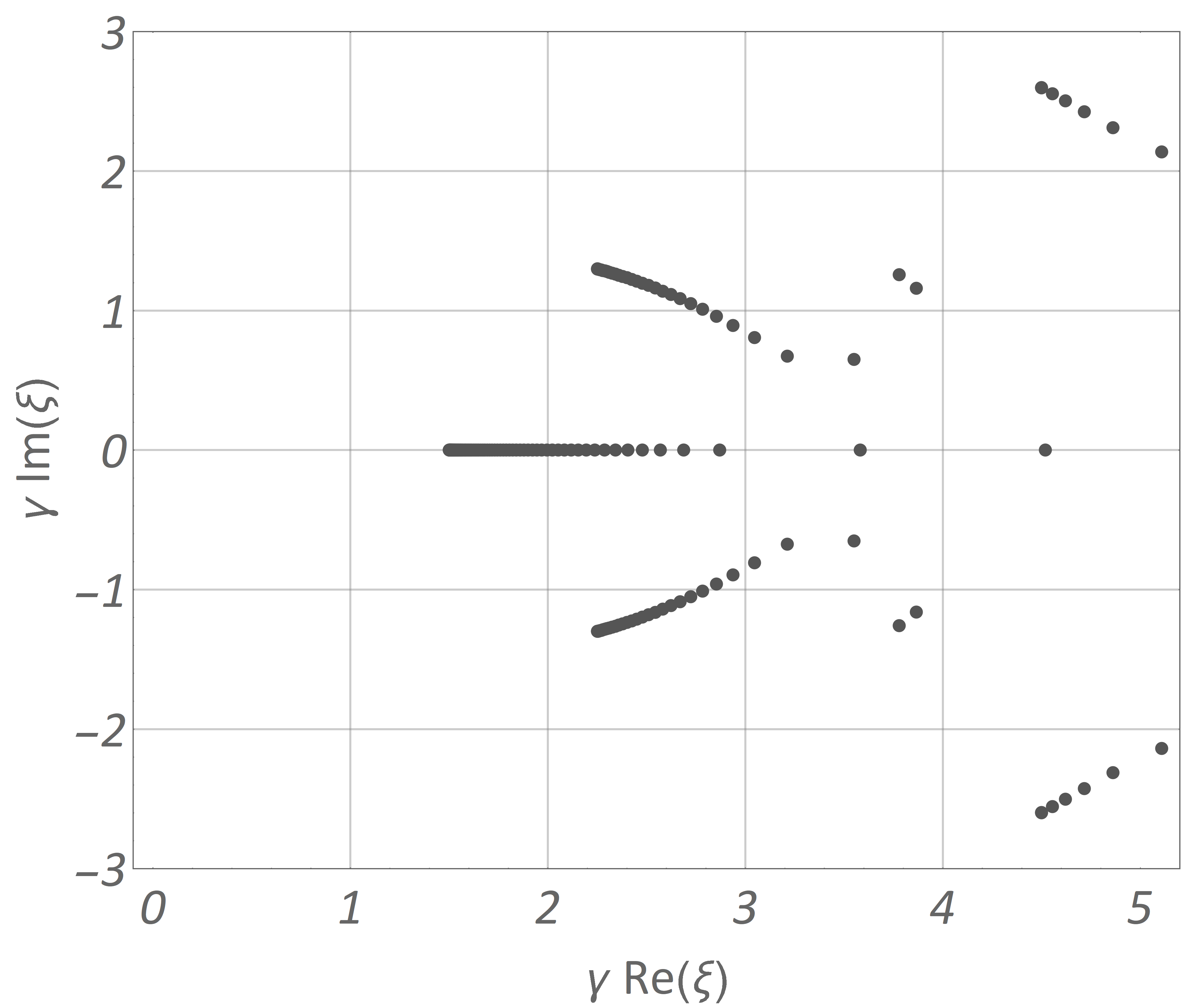}
\caption{Poles of the Pad{\'e} approximant to the Borel transform of the
  gradient expansion.
  We have checked that the structure of singularities remains stable as the number of terms kept in the series is varied. We have also checked that the residues of the pictured poles lie well above what was set as the numerical accuracy, i.e. that they are not numerical artefacts. The depicted singularities are discretizations of branch cuts with branch points at values of $\xi$ given in Eqs.~\eqref{eq.leadsing} and~\eqref{eq.omegacomplex}, as well as at $\xi_{0} + \xi_{\pm}$ and $2\, \xi_{\pm}$ (see also Ref.~\cite{Casalderrey-Solana:2017zyh} for a related statement based on a smaller data set from the previous version of the present manuscript).
}
\label{fig:poles}
\end{figure}
The Borel transform removes the leading order factorial growth of the
coefficients
\bel{borel}
\pa_B(\xi) = \sum_{n=1}^\infty \frac{\pac_n}{n!} \xi^{n}.
\ee
The inverse transform is given by the Borel summation formula
\bel{iborel}
\pa_{{\mathrm{resummed}}}(w) = \frac{1}{w} \int_{0}^{\infty} d\xi \, e^{-\xi/w} \, \pa_{B}(\xi)
\ee
and is not uniquely defined since the analytic continuation of $\pa_B(\xi)$
necessarily contains singularities that are responsible for the vanishing
radius of convergence of the original series.
We analytically continue the series from Eq.~\eqref{borel} truncated at 426 terms by means of Pad{\'e} approximants.
Fig.~\ref{fig:poles} shows the poles of the Pad{\'e} approximant, which condense in a way known to signify a branch-point singularity \cite{PadeCut}.

In Ref.~\cite{Heller:2015dha} (see also Refs.~\cite{Heller:2013fn,Basar:2015ava,Aniceto:2015mto,Buchel:2016cbj,Florkowski:2016zsi})
the ambiguity in the Borel summation associated
with singularities of the analytic continuation of the Borel transform has
been argued to disappear once the gradient expansion is supplemented with
exponentially decaying terms
\bel{expon}
\delta \pa \sim e^{-\xi_{0} \,  w},
\ee
where $\xi_{0}$ denotes the beginning of a cut in the complex Borel plane. In
the case under consideration the cut closest to the origin starts, up to 5 decimal places, at
\be
\label{eq.leadsing}
\xi_{0} = 1.5000/\gamma.
\ee

The constants $\xi_{0}$ appearing in \rf{expon} can in general be complex
(they then come in conjugate pairs) and in the examples analyzed so far in the
literature they appear in positive integer multiples. Each such term comes with
an infinite gradient
expansion of its own and the term with the lowest $\xi_{0}$ in a given family
comes with an independent complex integration constant. In all known
cases~\cite{Heller:2013fn,Heller:2015dha,Basar:2015ava,Aniceto:2015mto,Buchel:2016cbj} (see Ref.~\cite{Florkowski:2017olj} for a review),
those least damped modes within a given family coincide with singularities
of retarded equilibrium 2-point functions of the energy-momentum tensor at
vanishing momentum.

In the context of the RTA kinetic theory, the studies of
Ref.~\cite{Romatschke:2015gic} reveal the presence of a zero-momentum
fast-evolving mode in the isotropization of the energy-momentum tensor to its
equilibrium form:
\be
\label{eq.RTAqnm}
\delta \langle T^{\mu \nu} \rangle \sim e^{- \omega_{0} T \, t}, \quad
\omega_{0} = 1/\gamma.
\ee
The nontrivial background flow is known to modify the above equation to the form~\cite{Janik:2006gp,Heller:2013fn,Heller:2014wfa}
\be
\delta \langle T^{\mu \nu} \rangle \sim e^{- \omega_{0} \int T(x) \, u_{\mu} \, dx^{\mu}},
\ee
which for the case of Bjorken flow, neglecting subleading terms at large
values of
$w$, reduces to
\be
\label{eq.match}
\delta \pa \sim e^{-\frac{3}{2} \omega_{0} w}.
\ee
This, together with Eq.~\eqref{eq.RTAqnm}, reproduces Eq.~\eqref{eq.leadsing}.
Let us also note here that the analysis in Ref.~\cite{Heller:2018qvh} reveals that the cut along the real axis seen in Fig.~\ref{fig:poles} must be, in fact, an infinite collection of independent cuts. They all start at the same branch point, i.e. $\xi = \xi_{0}$, but are characterized by different discontinuities and are interpreted as an infinite set of modes carrying information about the initial distribution function to late times $w$.

Importantly, Fig.~\ref{fig:poles} contains also a pair of
singularities characterized by
\be
\label{eq.omegacomplex}
\xi_{\pm} \approx (2.25016 \pm 1.29898 \, i)/\gamma .
\ee
If one were to apply straightforwardly the lessons from earlier studies of other models of expanding plasmas, these
complex values of $\xi_{0}$ would be interpreted as oscillatory-type transient contributions to the pressure anisotropy.
As shown in the next section, where we argue that these are unphysical, this natural-looking conclusion is premature. This is an important point, since such
singularities appear also in other kinetic theory models, such as those with $\reltime\sim T^{-\Delta}$, for $0 < \Delta < 3$. Notably, for $\Delta>2$, the unphysical modes are actually closest to the origin, so naively they would correspond to the dominant nonhydrodynamic corrections~\cite{Heller:2018qvh}.

\vspace{10 pt}

\noindent \emph{5. Analytic properties of RTA kinetic theory--}
In order to
explain the singularities of the Borel transform at $\xi_{\pm}$, see Eq.~\eqref{eq.omegacomplex} and
Fig.~\ref{fig:poles}, we use the integral equation~\cite{Florkowski:2013lya} which follows from the Boltzmann
equation~\cite{Baym:1984np} and directly determines the local energy density ${\cal E}(\tau)$
\be
\label{eq.integral}
g(\tau) = {\cal E}_{0} (\tau) + \frac{1}{2} \, \int_{\tau_{0}}^{\tau} \frac{\mathrm{d}\tau'}{\reltime(\tau')} \, H\left( \frac{\tau'}{\tau}\right) \, g(\tau').
\ee
In the expression above, ${\cal E}_{0}(\tau)$ carries information about initial conditions, but will not be relevant in the following analysis. The object of interest is the energy density ${\cal E}(\tau)$, which~appears~in
\be
\label{eq.defg}
g(\tau) = {\cal E}(\tau) \, e^{\int_{\tau_{0}}^{\tau} \frac{\mathrm{d}\tau'}{\reltime(\tau')}},
\ee
as well as in $\reltime(\tau)$ through Eqs.~\eqref{eq.reltime} and~\eqref{lmc}. The function $H(q)$ originates from the second moment of the equilibrium distribution function, Eq.~\eqref{boldist}, and reads
\be
H(q) = q^{2} + \frac{\arctan{\sqrt{\frac{1}{q^{2}}-1}}}{\sqrt{\frac{1}{q^{2}}-1}}.
\ee
What will be crucial in the following is the analytic structure of $H$.
Since the natural variable in our considerations is $w$ and at late times $w \sim \tau^{2/3}$,
	we shall write the argument of $H$ as $q = (w'/w)^{3/2} \equiv \zeta^{3/2}$.

Among the singularities of $H\left(\zeta^{3/2}\right)$, the one of interest is the  branch point stemming from the square root in the denominator. It appears as a singularity when the $\arctan$ function in the numerator is taken in a non-principal branch.
As a result, one finds
\be
\label{eq.Hsing}
H\left(\zeta^{3/2}\right) \sim \left(1 - \zeta^{3}\right)^{-1/2}.
\ee
which has branch points at third roots of unity. We will be interested in the ones located at
\be
\zeta_{\pm} = e^{\pm \, i \, \frac{2}{3} \, \pi}.
\ee
The key observation is that the presence of singularities in the complex $\zeta$ plane leads to inequivalent choices of integration contours in Eq.~\eqref{eq.integral}, the only physical choice being homologous to the integration along the real axis. If one uses the late time solution for $g(\tau(w))$ obtained from Eq.~\eqref{eq.r123} under the integral in Eq.~\eqref{eq.integral} and considers two inequivalent contours around $\zeta_{-}$ (or, similarly, $\zeta_{+}$), denoted ${\cal C}_{1}$ and ${\cal C}_{2}$, one finds
\be
\delta g(w) \sim \left[ \int_{{\cal C}_{1}} \mathrm{d} \zeta - \int_{{\cal C}_{2}} \mathrm{d} \zeta \right] \, e^{\frac{3 \, w}{2\,\gamma}  \, \zeta}  \, H(\zeta^{3/2}) \times \ldots,
\ee
where the elipsis denotes terms subleading at large $w$.
Similarly to the analysis around Eq.~\eqref{iborel}, the branch cuts lead to contributions to $\delta g$ of the form
\be
\delta g \sim e^{- \frac{3\zeta_{\pm}}{2\,\gamma} \, w},
\ee
where we truncated subleading terms at large values of~$w$. Finally, note that $\delta g$ and $\delta {\cal E}$ are related through Eq.~\eqref{eq.defg} with, at late times / large values of $w$, $e^{\int_{\tau_{0}}^{\tau} \frac{\mathrm{d}\tau'}{\reltime(\tau')}} \sim e^{\frac{3}{2\, \gamma} \, w}$ which ultimately gives
\be
\delta {\cal E}_{\pm} \sim e^{-\frac{3}{2\,\gamma} \, \left( \zeta_{\pm} - 1 \right) \, w}.
\ee
Comparing the exponent in the equation above with Eq.~\eqref{eq.omegacomplex} one observes a remarkable agreement up to 4 decimal places. As a further way of corroborating this result, one can use the techniques utilized earlier in this context in Refs.~\cite{Heller:2013fn,Heller:2015dha} to match the square root character of the branch cut in Eq.~\eqref{eq.Hsing} with the leading power-law correction in $w$ to contributions from the $\xi_{\pm}$ singularities in the Borel plane, with very good agreement.  

All this taken together gives us confidence that the correct interpretation of the singularities
given in Eq.~\eqref{eq.omegacomplex} is that they correspond not to physical excitations,ut rather to analytic
properties of kinetic theory for complexified values of the $w$ variable. The physical integration contour in
Eq.~\eqref{eq.integral} along real values of $\tau'$ does not pick up these contributions (and, no wonder, they are not
seen in solutions of the initial value problem discussed in Ref.~\cite{Heller:2018qvh}). The gradient expansion
itself does not preclude unphysical choices of contour, and this is reflected in its large-order behaviour.  
Similar phenomena can be seen in the integral equation considered in Ref.~\cite{Aniceto:2015rua} or in the ghost-instanton story of Ref. \cite{Basar:2013eka}. Their origin goes back to the fundamental point  of resurgence: all non-perturbative information is encoded in the large-order behavior of the perturbative sector. 

\vspace{10 pt}
\noindent \emph{6. Hydrodynamization compared--}
In previous sections we have demonstrated that hydrodynamization in the RTA takes place through the decay of transient nonhydrodynamical modes in complete analogy to the situation at strong coupling. On the other hand, the RTA shares structural similarities with EKT -- while the collision kernels of the QCD effective kinetic theory have much richer structure than the RTA, many qualitative features at the level of kinetic theory coincide. 
In this section we further strengthen the connection between RTA and EKT by noting that the similarity between these two theories goes
beyond abstract structural similarity and that they agree semi-quantitatively
when the values of $\eta/s$ are matched between the two theories.

To demonstrate the extent of quantitative agreement between different models,
inspired by Ref.~\cite{Keegan:2015avk} we introduce a new, rescaled variable $\tilde{w}\equiv \frac{w}{4\pi \eta/s}$.
This is useful for such
comparisons, since the late time behaviour of the pressure anisotropy $\pa$ is given by
$$
\pa_H(\tilde w) = \frac{2}{\pi \tilde w} + \mathcal{O}\left(\frac{1}{\tilde w^2}\right).
$$
The leading behaviour is completely universal and does not depend on the value
of $\eta/s$. Deviations from the asymptotic form characterize contributions
arising beyond first order hydrodynamics and
indeed we say that the system has reached the hydrodynamic regime when for a given state the relative difference between $\pa$ and $\pa_H$ remains smaller than some threshold value.
\rff{fig:tilde} shows a comparison of the time evolution of the system evolved
in the EKT from \cite{Kurkela:2015qoa}, RTA using the methodology of
\cite{Baym:1984np,Florkowski:2013lya}, and numerical AdS/CFT calculation of
\cite{Heller:2011ju,Heller:2012je,Jankowski:2014lna}. For the EKT and RTA
simulations we took the initial condition used in \cite{Kurkela:2015qoa},
whereas for the AdS/CFT simulation we took a typical initial conditions from
\cite{Jankowski:2014lna}. We evolved the systems using EKT with $\lambda =
10$ corresponding to $\eta/s \approx 0.642$, holography with $\eta/s = 1/4\pi$
and RTA with $\gamma$ fixed to reproduce the value of $\eta/s$ of either
model. Note that every kinetic theory curve in \rff{fig:tilde} corresponds to a
different initial distribution function.
In all these models the evolution is similar, but
distinct. Remarkably, in each case -- despite vastly differing microphysics --
the evolution converges to first order viscous hydrodynamics roughly at same
$\tilde w \approx 1$. It is striking that in all these cases the pressure
anisotropy at the time of hydrodynamization is as high as $\pa\approx
0.6-0.8$. 

The structural and quantitative similarities of EKT and RTA suggests that the gradient expansion in EKT also exhibits zero radius of convergence and that the weak coupling hydrodynamization is driven by the same qualitative process. This is connected with the notion of attractors which have been explored in both RTA and holography~\cite{Romatschke:2017vte,Spalinski:2017mel}. To what extent these insights translate to EKT is an important problem.

\vspace{10 pt} \noindent \emph{7. Acknowledgments--} We would like to thank the authors of \rfc{Bazow:2015zca} as well as I. Aniceto, G. Dunne, W.~Florkowski, R.~Janik,
P.~Romatschke, W. van der Schee, M. Martinez and M.~Strickland for valuable discussions and correspondence. Research at Perimeter
Institute is supported by the Government of Canada through Industry Canada and by the Province of Ontario through the
Ministry of Research \& Innovation. M.P.H. acknowledges support from the Alexander von Humboldt Foundation and the
Federal Ministry for Education and Research through the Sofja Kovalevskaja Award. M.S. and V.S were supported by the
Polish National Science Centre Grant 2015/19/B/ST2/02824.

\bibliography{hydrok}{}

\begin{thebibliography}{60}%
\makeatletter
\providecommand \@ifxundefined [1]{%
 \@ifx{#1\undefined}
}%
\providecommand \@ifnum [1]{%
 \ifnum #1\expandafter \@firstoftwo
 \else \expandafter \@secondoftwo
 \fi
}%
\providecommand \@ifx [1]{%
 \ifx #1\expandafter \@firstoftwo
 \else \expandafter \@secondoftwo
 \fi
}%
\providecommand \natexlab [1]{#1}%
\providecommand \enquote  [1]{``#1''}%
\providecommand \bibnamefont  [1]{#1}%
\providecommand \bibfnamefont [1]{#1}%
\providecommand \citenamefont [1]{#1}%
\providecommand \href@noop [0]{\@secondoftwo}%
\providecommand \href [0]{\begingroup \@sanitize@url \@href}%
\providecommand \@href[1]{\@@startlink{#1}\@@href}%
\providecommand \@@href[1]{\endgroup#1\@@endlink}%
\providecommand \@sanitize@url [0]{\catcode `\\12\catcode `\$12\catcode
  `\&12\catcode `\#12\catcode `\^12\catcode `\_12\catcode `\%12\relax}%
\providecommand \@@startlink[1]{}%
\providecommand \@@endlink[0]{}%
\providecommand \url  [0]{\begingroup\@sanitize@url \@url }%
\providecommand \@url [1]{\endgroup\@href {#1}{\urlprefix }}%
\providecommand \urlprefix  [0]{URL }%
\providecommand \Eprint [0]{\href }%
\providecommand \doibase [0]{http://dx.doi.org/}%
\providecommand \selectlanguage [0]{\@gobble}%
\providecommand \bibinfo  [0]{\@secondoftwo}%
\providecommand \bibfield  [0]{\@secondoftwo}%
\providecommand \translation [1]{[#1]}%
\providecommand \BibitemOpen [0]{}%
\providecommand \bibitemStop [0]{}%
\providecommand \bibitemNoStop [0]{.\EOS\space}%
\providecommand \EOS [0]{\spacefactor3000\relax}%
\providecommand \BibitemShut  [1]{\csname bibitem#1\endcsname}%
\let\auto@bib@innerbib\@empty
\bibitem [{\citenamefont {Florkowski}\ \emph {et~al.}(2018)\citenamefont
  {Florkowski}, \citenamefont {Heller},\ and\ \citenamefont
  {Spalinski}}]{Florkowski:2017olj}%
  \BibitemOpen
  \bibfield  {author} {\bibinfo {author} {\bibfnamefont {W.}~\bibnamefont
  {Florkowski}}, \bibinfo {author} {\bibfnamefont {M.~P.}\ \bibnamefont
  {Heller}}, \ and\ \bibinfo {author} {\bibfnamefont {M.}~\bibnamefont
  {Spalinski}},\ }\href {\doibase 10.1088/1361-6633/aaa091} {\bibfield
  {journal} {\bibinfo  {journal} {Rept. Prog. Phys.}\ }\textbf {\bibinfo
  {volume} {81}},\ \bibinfo {pages} {046001} (\bibinfo {year} {2018})},\
  \Eprint {http://arxiv.org/abs/1707.02282} {arXiv:1707.02282 [hep-ph]}
  \BibitemShut {NoStop}%
\bibitem [{\citenamefont {Romatschke}\ and\ \citenamefont
  {Romatschke}(2017)}]{Romatschke:2017ejr}%
  \BibitemOpen
  \bibfield  {author} {\bibinfo {author} {\bibfnamefont {P.}~\bibnamefont
  {Romatschke}}\ and\ \bibinfo {author} {\bibfnamefont {U.}~\bibnamefont
  {Romatschke}},\ }\href@noop {} {\  (\bibinfo {year} {2017})},\ \Eprint
  {http://arxiv.org/abs/1712.05815} {arXiv:1712.05815 [nucl-th]} \BibitemShut
  {NoStop}%
\bibitem [{\citenamefont {Arnold}\ \emph {et~al.}(2003)\citenamefont {Arnold},
  \citenamefont {Moore},\ and\ \citenamefont {Yaffe}}]{Arnold:2002zm}%
  \BibitemOpen
  \bibfield  {author} {\bibinfo {author} {\bibfnamefont {P.~B.}\ \bibnamefont
  {Arnold}}, \bibinfo {author} {\bibfnamefont {G.~D.}\ \bibnamefont {Moore}}, \
  and\ \bibinfo {author} {\bibfnamefont {L.~G.}\ \bibnamefont {Yaffe}},\ }\href
  {\doibase 10.1088/1126-6708/2003/01/030} {\bibfield  {journal} {\bibinfo
  {journal} {JHEP}\ }\textbf {\bibinfo {volume} {01}},\ \bibinfo {pages} {030}
  (\bibinfo {year} {2003})},\ \Eprint {http://arxiv.org/abs/hep-ph/0209353}
  {arXiv:hep-ph/0209353 [hep-ph]} \BibitemShut {NoStop}%
\bibitem [{\citenamefont {Baier}\ \emph {et~al.}(2001)\citenamefont {Baier},
  \citenamefont {Mueller}, \citenamefont {Schiff},\ and\ \citenamefont
  {Son}}]{Baier:2000sb}%
  \BibitemOpen
  \bibfield  {author} {\bibinfo {author} {\bibfnamefont {R.}~\bibnamefont
  {Baier}}, \bibinfo {author} {\bibfnamefont {A.~H.}\ \bibnamefont {Mueller}},
  \bibinfo {author} {\bibfnamefont {D.}~\bibnamefont {Schiff}}, \ and\ \bibinfo
  {author} {\bibfnamefont {D.~T.}\ \bibnamefont {Son}},\ }\href {\doibase
  10.1016/S0370-2693(01)00191-5} {\bibfield  {journal} {\bibinfo  {journal}
  {Phys. Lett.}\ }\textbf {\bibinfo {volume} {B502}},\ \bibinfo {pages} {51}
  (\bibinfo {year} {2001})},\ \Eprint {http://arxiv.org/abs/hep-ph/0009237}
  {arXiv:hep-ph/0009237 [hep-ph]} \BibitemShut {NoStop}%
\bibitem [{\citenamefont {Arnold}\ \emph {et~al.}(2006)\citenamefont {Arnold},
  \citenamefont {Dogan},\ and\ \citenamefont {Moore}}]{Arnold:2006fz}%
  \BibitemOpen
  \bibfield  {author} {\bibinfo {author} {\bibfnamefont {P.~B.}\ \bibnamefont
  {Arnold}}, \bibinfo {author} {\bibfnamefont {C.}~\bibnamefont {Dogan}}, \
  and\ \bibinfo {author} {\bibfnamefont {G.~D.}\ \bibnamefont {Moore}},\ }\href
  {\doibase 10.1103/PhysRevD.74.085021} {\bibfield  {journal} {\bibinfo
  {journal} {Phys. Rev.}\ }\textbf {\bibinfo {volume} {D74}},\ \bibinfo {pages}
  {085021} (\bibinfo {year} {2006})},\ \Eprint
  {http://arxiv.org/abs/hep-ph/0608012} {arXiv:hep-ph/0608012 [hep-ph]}
  \BibitemShut {NoStop}%
\bibitem [{\citenamefont {Kurkela}\ and\ \citenamefont
  {Moore}(2011{\natexlab{a}})}]{Kurkela:2011ti}%
  \BibitemOpen
  \bibfield  {author} {\bibinfo {author} {\bibfnamefont {A.}~\bibnamefont
  {Kurkela}}\ and\ \bibinfo {author} {\bibfnamefont {G.~D.}\ \bibnamefont
  {Moore}},\ }\href {\doibase 10.1007/JHEP12(2011)044} {\bibfield  {journal}
  {\bibinfo  {journal} {JHEP}\ }\textbf {\bibinfo {volume} {12}},\ \bibinfo
  {pages} {044} (\bibinfo {year} {2011}{\natexlab{a}})},\ \Eprint
  {http://arxiv.org/abs/1107.5050} {arXiv:1107.5050 [hep-ph]} \BibitemShut
  {NoStop}%
\bibitem [{\citenamefont {Kurkela}\ and\ \citenamefont
  {Moore}(2011{\natexlab{b}})}]{Kurkela:2011ub}%
  \BibitemOpen
  \bibfield  {author} {\bibinfo {author} {\bibfnamefont {A.}~\bibnamefont
  {Kurkela}}\ and\ \bibinfo {author} {\bibfnamefont {G.~D.}\ \bibnamefont
  {Moore}},\ }\href {\doibase 10.1007/JHEP11(2011)120} {\bibfield  {journal}
  {\bibinfo  {journal} {JHEP}\ }\textbf {\bibinfo {volume} {11}},\ \bibinfo
  {pages} {120} (\bibinfo {year} {2011}{\natexlab{b}})},\ \Eprint
  {http://arxiv.org/abs/1108.4684} {arXiv:1108.4684 [hep-ph]} \BibitemShut
  {NoStop}%
\bibitem [{\citenamefont {Kurkela}\ and\ \citenamefont
  {Lu}(2014)}]{Kurkela:2014tea}%
  \BibitemOpen
  \bibfield  {author} {\bibinfo {author} {\bibfnamefont {A.}~\bibnamefont
  {Kurkela}}\ and\ \bibinfo {author} {\bibfnamefont {E.}~\bibnamefont {Lu}},\
  }\href {\doibase 10.1103/PhysRevLett.113.182301} {\bibfield  {journal}
  {\bibinfo  {journal} {Phys. Rev. Lett.}\ }\textbf {\bibinfo {volume} {113}},\
  \bibinfo {pages} {182301} (\bibinfo {year} {2014})},\ \Eprint
  {http://arxiv.org/abs/1405.6318} {arXiv:1405.6318 [hep-ph]} \BibitemShut
  {NoStop}%
\bibitem [{\citenamefont {Kurkela}\ and\ \citenamefont
  {Zhu}(2015)}]{Kurkela:2015qoa}%
  \BibitemOpen
  \bibfield  {author} {\bibinfo {author} {\bibfnamefont {A.}~\bibnamefont
  {Kurkela}}\ and\ \bibinfo {author} {\bibfnamefont {Y.}~\bibnamefont {Zhu}},\
  }\href {\doibase 10.1103/PhysRevLett.115.182301} {\bibfield  {journal}
  {\bibinfo  {journal} {Phys. Rev. Lett.}\ }\textbf {\bibinfo {volume} {115}},\
  \bibinfo {pages} {182301} (\bibinfo {year} {2015})},\ \Eprint
  {http://arxiv.org/abs/1506.06647} {arXiv:1506.06647 [hep-ph]} \BibitemShut
  {NoStop}%
\bibitem [{\citenamefont {Maldacena}(1998)}]{Maldacena:1997re}%
  \BibitemOpen
  \bibfield  {author} {\bibinfo {author} {\bibfnamefont {J.~M.}\ \bibnamefont
  {Maldacena}},\ }\href@noop {} {\bibfield  {journal} {\bibinfo  {journal}
  {Adv.Theor.Math.Phys.}\ }\textbf {\bibinfo {volume} {2}},\ \bibinfo {pages}
  {231} (\bibinfo {year} {1998})},\ \Eprint
  {http://arxiv.org/abs/hep-th/9711200} {arXiv:hep-th/9711200 [hep-th]}
  \BibitemShut {NoStop}%
\bibitem [{\citenamefont {Chesler}\ and\ \citenamefont
  {Yaffe}(2010)}]{Chesler:2009cy}%
  \BibitemOpen
  \bibfield  {author} {\bibinfo {author} {\bibfnamefont {P.~M.}\ \bibnamefont
  {Chesler}}\ and\ \bibinfo {author} {\bibfnamefont {L.~G.}\ \bibnamefont
  {Yaffe}},\ }\href {\doibase 10.1103/PhysRevD.82.026006} {\bibfield  {journal}
  {\bibinfo  {journal} {Phys.Rev.}\ }\textbf {\bibinfo {volume} {D82}},\
  \bibinfo {pages} {026006} (\bibinfo {year} {2010})},\ \Eprint
  {http://arxiv.org/abs/0906.4426} {arXiv:0906.4426 [hep-th]} \BibitemShut
  {NoStop}%
\bibitem [{\citenamefont {Chesler}\ and\ \citenamefont
  {Yaffe}(2011)}]{Chesler:2010bi}%
  \BibitemOpen
  \bibfield  {author} {\bibinfo {author} {\bibfnamefont {P.~M.}\ \bibnamefont
  {Chesler}}\ and\ \bibinfo {author} {\bibfnamefont {L.~G.}\ \bibnamefont
  {Yaffe}},\ }\href {\doibase 10.1103/PhysRevLett.106.021601} {\bibfield
  {journal} {\bibinfo  {journal} {Phys.Rev.Lett.}\ }\textbf {\bibinfo {volume}
  {106}},\ \bibinfo {pages} {021601} (\bibinfo {year} {2011})},\ \Eprint
  {http://arxiv.org/abs/1011.3562} {arXiv:1011.3562 [hep-th]} \BibitemShut
  {NoStop}%
\bibitem [{\citenamefont {Heller}\ \emph
  {et~al.}(2012{\natexlab{a}})\citenamefont {Heller}, \citenamefont {Janik},\
  and\ \citenamefont {Witaszczyk}}]{Heller:2011ju}%
  \BibitemOpen
  \bibfield  {author} {\bibinfo {author} {\bibfnamefont {M.~P.}\ \bibnamefont
  {Heller}}, \bibinfo {author} {\bibfnamefont {R.~A.}\ \bibnamefont {Janik}}, \
  and\ \bibinfo {author} {\bibfnamefont {P.}~\bibnamefont {Witaszczyk}},\
  }\href {\doibase 10.1103/PhysRevLett.108.201602} {\bibfield  {journal}
  {\bibinfo  {journal} {Phys.Rev.Lett.}\ }\textbf {\bibinfo {volume} {108}},\
  \bibinfo {pages} {201602} (\bibinfo {year} {2012}{\natexlab{a}})},\ \Eprint
  {http://arxiv.org/abs/1103.3452} {arXiv:1103.3452 [hep-th]} \BibitemShut
  {NoStop}%
\bibitem [{\citenamefont {Casalderrey-Solana}\ \emph
  {et~al.}(2011)\citenamefont {Casalderrey-Solana}, \citenamefont {Liu},
  \citenamefont {Mateos}, \citenamefont {Rajagopal},\ and\ \citenamefont
  {Wiedemann}}]{CasalderreySolana:2011us}%
  \BibitemOpen
  \bibfield  {author} {\bibinfo {author} {\bibfnamefont {J.}~\bibnamefont
  {Casalderrey-Solana}}, \bibinfo {author} {\bibfnamefont {H.}~\bibnamefont
  {Liu}}, \bibinfo {author} {\bibfnamefont {D.}~\bibnamefont {Mateos}},
  \bibinfo {author} {\bibfnamefont {K.}~\bibnamefont {Rajagopal}}, \ and\
  \bibinfo {author} {\bibfnamefont {U.~A.}\ \bibnamefont {Wiedemann}},\
  }\href@noop {} {\  (\bibinfo {year} {2011})},\ \Eprint
  {http://arxiv.org/abs/1101.0618} {arXiv:1101.0618 [hep-th]} \BibitemShut
  {NoStop}%
\bibitem [{\citenamefont {Heller}\ \emph {et~al.}(2013)\citenamefont {Heller},
  \citenamefont {Janik},\ and\ \citenamefont {Witaszczyk}}]{Heller:2013fn}%
  \BibitemOpen
  \bibfield  {author} {\bibinfo {author} {\bibfnamefont {M.~P.}\ \bibnamefont
  {Heller}}, \bibinfo {author} {\bibfnamefont {R.~A.}\ \bibnamefont {Janik}}, \
  and\ \bibinfo {author} {\bibfnamefont {P.}~\bibnamefont {Witaszczyk}},\
  }\href {\doibase 10.1103/PhysRevLett.110.211602} {\bibfield  {journal}
  {\bibinfo  {journal} {Phys.Rev.Lett.}\ }\textbf {\bibinfo {volume} {110}},\
  \bibinfo {pages} {211602} (\bibinfo {year} {2013})},\ \Eprint
  {http://arxiv.org/abs/1302.0697} {arXiv:1302.0697 [hep-th]} \BibitemShut
  {NoStop}%
\bibitem [{\citenamefont {Anderson}\ and\ \citenamefont
  {Witting}(1974)}]{Anderson}%
  \BibitemOpen
  \bibfield  {author} {\bibinfo {author} {\bibfnamefont {J.}~\bibnamefont
  {Anderson}}\ and\ \bibinfo {author} {\bibfnamefont {H.}~\bibnamefont
  {Witting}},\ }\href@noop {} {\bibfield  {journal} {\bibinfo  {journal}
  {Physica}\ }\textbf {\bibinfo {volume} {74}},\ \bibinfo {pages} {466}
  (\bibinfo {year} {1974})}\BibitemShut {NoStop}%
\bibitem [{\citenamefont {Buchel}\ \emph {et~al.}(2015)\citenamefont {Buchel},
  \citenamefont {Heller},\ and\ \citenamefont {Myers}}]{Buchel:2015saa}%
  \BibitemOpen
  \bibfield  {author} {\bibinfo {author} {\bibfnamefont {A.}~\bibnamefont
  {Buchel}}, \bibinfo {author} {\bibfnamefont {M.~P.}\ \bibnamefont {Heller}},
  \ and\ \bibinfo {author} {\bibfnamefont {R.~C.}\ \bibnamefont {Myers}},\
  }\href {\doibase 10.1103/PhysRevLett.114.251601} {\bibfield  {journal}
  {\bibinfo  {journal} {Phys. Rev. Lett.}\ }\textbf {\bibinfo {volume} {114}},\
  \bibinfo {pages} {251601} (\bibinfo {year} {2015})},\ \Eprint
  {http://arxiv.org/abs/1503.07114} {arXiv:1503.07114 [hep-th]} \BibitemShut
  {NoStop}%
\bibitem [{\citenamefont {Janik}\ \emph {et~al.}(2015)\citenamefont {Janik},
  \citenamefont {Plewa}, \citenamefont {Soltanpanahi},\ and\ \citenamefont
  {Spalinski}}]{Janik:2015waa}%
  \BibitemOpen
  \bibfield  {author} {\bibinfo {author} {\bibfnamefont {R.~A.}\ \bibnamefont
  {Janik}}, \bibinfo {author} {\bibfnamefont {G.}~\bibnamefont {Plewa}},
  \bibinfo {author} {\bibfnamefont {H.}~\bibnamefont {Soltanpanahi}}, \ and\
  \bibinfo {author} {\bibfnamefont {M.}~\bibnamefont {Spalinski}},\ }\href
  {\doibase 10.1103/PhysRevD.91.126013} {\bibfield  {journal} {\bibinfo
  {journal} {Phys. Rev.}\ }\textbf {\bibinfo {volume} {D91}},\ \bibinfo {pages}
  {126013} (\bibinfo {year} {2015})},\ \Eprint
  {http://arxiv.org/abs/1503.07149} {arXiv:1503.07149 [hep-th]} \BibitemShut
  {NoStop}%
\bibitem [{\citenamefont {Denicol}\ and\ \citenamefont
  {Noronha}(2016)}]{Denicol:2016bjh}%
  \BibitemOpen
  \bibfield  {author} {\bibinfo {author} {\bibfnamefont {G.~S.}\ \bibnamefont
  {Denicol}}\ and\ \bibinfo {author} {\bibfnamefont {J.}~\bibnamefont
  {Noronha}},\ }\href@noop {} {\  (\bibinfo {year} {2016})},\ \Eprint
  {http://arxiv.org/abs/1608.07869} {arXiv:1608.07869 [nucl-th]} \BibitemShut
  {NoStop}%
\bibitem [{\citenamefont {Baier}\ \emph {et~al.}(2008)\citenamefont {Baier},
  \citenamefont {Romatschke}, \citenamefont {Son}, \citenamefont {Starinets},\
  and\ \citenamefont {Stephanov}}]{Baier:2007ix}%
  \BibitemOpen
  \bibfield  {author} {\bibinfo {author} {\bibfnamefont {R.}~\bibnamefont
  {Baier}}, \bibinfo {author} {\bibfnamefont {P.}~\bibnamefont {Romatschke}},
  \bibinfo {author} {\bibfnamefont {D.~T.}\ \bibnamefont {Son}}, \bibinfo
  {author} {\bibfnamefont {A.~O.}\ \bibnamefont {Starinets}}, \ and\ \bibinfo
  {author} {\bibfnamefont {M.~A.}\ \bibnamefont {Stephanov}},\ }\href {\doibase
  10.1088/1126-6708/2008/04/100} {\bibfield  {journal} {\bibinfo  {journal}
  {JHEP}\ }\textbf {\bibinfo {volume} {04}},\ \bibinfo {pages} {100} (\bibinfo
  {year} {2008})},\ \Eprint {http://arxiv.org/abs/0712.2451} {arXiv:0712.2451
  [hep-th]} \BibitemShut {NoStop}%
\bibitem [{\citenamefont {Heller}\ and\ \citenamefont
  {Spalinski}(2015)}]{Heller:2015dha}%
  \BibitemOpen
  \bibfield  {author} {\bibinfo {author} {\bibfnamefont {M.~P.}\ \bibnamefont
  {Heller}}\ and\ \bibinfo {author} {\bibfnamefont {M.}~\bibnamefont
  {Spalinski}},\ }\href {\doibase 10.1103/PhysRevLett.115.072501} {\bibfield
  {journal} {\bibinfo  {journal} {Phys. Rev. Lett.}\ }\textbf {\bibinfo
  {volume} {115}},\ \bibinfo {pages} {072501} (\bibinfo {year} {2015})},\
  \Eprint {http://arxiv.org/abs/1503.07514} {arXiv:1503.07514 [hep-th]}
  \BibitemShut {NoStop}%
\bibitem [{\citenamefont {Romatschke}(2016)}]{Romatschke:2015gic}%
  \BibitemOpen
  \bibfield  {author} {\bibinfo {author} {\bibfnamefont {P.}~\bibnamefont
  {Romatschke}},\ }\href {\doibase 10.1140/epjc/s10052-016-4169-7} {\bibfield
  {journal} {\bibinfo  {journal} {Eur. Phys. J.}\ }\textbf {\bibinfo {volume}
  {C76}},\ \bibinfo {pages} {352} (\bibinfo {year} {2016})},\ \Eprint
  {http://arxiv.org/abs/1512.02641} {arXiv:1512.02641 [hep-th]} \BibitemShut
  {NoStop}%
\bibitem [{\citenamefont {Kurkela}\ and\ \citenamefont
  {Wiedemann}(2017)}]{Kurkela:2017xis}%
  \BibitemOpen
  \bibfield  {author} {\bibinfo {author} {\bibfnamefont {A.}~\bibnamefont
  {Kurkela}}\ and\ \bibinfo {author} {\bibfnamefont {U.~A.}\ \bibnamefont
  {Wiedemann}},\ }\href@noop {} {\  (\bibinfo {year} {2017})},\ \Eprint
  {http://arxiv.org/abs/1712.04376} {arXiv:1712.04376 [hep-ph]} \BibitemShut
  {NoStop}%
\bibitem [{\citenamefont {Romatschke}(2015)}]{Romatschke:2015gxa}%
  \BibitemOpen
  \bibfield  {author} {\bibinfo {author} {\bibfnamefont {P.}~\bibnamefont
  {Romatschke}},\ }\href {\doibase 10.1140/epjc/s10052-015-3509-3} {\bibfield
  {journal} {\bibinfo  {journal} {Eur. Phys. J.}\ }\textbf {\bibinfo {volume}
  {C75}},\ \bibinfo {pages} {305} (\bibinfo {year} {2015})},\ \Eprint
  {http://arxiv.org/abs/1502.04745} {arXiv:1502.04745 [nucl-th]} \BibitemShut
  {NoStop}%
\bibitem [{\citenamefont {Habich}\ \emph {et~al.}(2015)\citenamefont {Habich},
  \citenamefont {Miller}, \citenamefont {Romatschke},\ and\ \citenamefont
  {Xiang}}]{Habich:2015rtj}%
  \BibitemOpen
  \bibfield  {author} {\bibinfo {author} {\bibfnamefont {M.}~\bibnamefont
  {Habich}}, \bibinfo {author} {\bibfnamefont {G.~A.}\ \bibnamefont {Miller}},
  \bibinfo {author} {\bibfnamefont {P.}~\bibnamefont {Romatschke}}, \ and\
  \bibinfo {author} {\bibfnamefont {W.}~\bibnamefont {Xiang}},\ }\href@noop {}
  {\  (\bibinfo {year} {2015})},\ \Eprint {http://arxiv.org/abs/1512.05354}
  {arXiv:1512.05354 [nucl-th]} \BibitemShut {NoStop}%
\bibitem [{\citenamefont {Spalinski}(2016)}]{Spalinski:2016fnj}%
  \BibitemOpen
  \bibfield  {author} {\bibinfo {author} {\bibfnamefont {M.}~\bibnamefont
  {Spalinski}},\ }\href {\doibase 10.1103/PhysRevD.94.085002} {\bibfield
  {journal} {\bibinfo  {journal} {Phys. Rev.}\ }\textbf {\bibinfo {volume}
  {D94}},\ \bibinfo {pages} {085002} (\bibinfo {year} {2016})},\ \Eprint
  {http://arxiv.org/abs/1607.06381} {arXiv:1607.06381 [nucl-th]} \BibitemShut
  {NoStop}%
\bibitem [{\citenamefont
  {Romatschke}(2017{\natexlab{a}})}]{Romatschke:2016hle}%
  \BibitemOpen
  \bibfield  {author} {\bibinfo {author} {\bibfnamefont {P.}~\bibnamefont
  {Romatschke}},\ }\href {\doibase 10.1140/epjc/s10052-016-4567-x} {\bibfield
  {journal} {\bibinfo  {journal} {Eur. Phys. J.}\ }\textbf {\bibinfo {volume}
  {C77}},\ \bibinfo {pages} {21} (\bibinfo {year} {2017}{\natexlab{a}})},\
  \Eprint {http://arxiv.org/abs/1609.02820} {arXiv:1609.02820 [nucl-th]}
  \BibitemShut {NoStop}%
\bibitem [{\citenamefont {Romatschke}(2018)}]{Romatschke:2017vte}%
  \BibitemOpen
  \bibfield  {author} {\bibinfo {author} {\bibfnamefont {P.}~\bibnamefont
  {Romatschke}},\ }\href {\doibase 10.1103/PhysRevLett.120.012301} {\bibfield
  {journal} {\bibinfo  {journal} {Phys. Rev. Lett.}\ }\textbf {\bibinfo
  {volume} {120}},\ \bibinfo {pages} {012301} (\bibinfo {year} {2018})},\
  \Eprint {http://arxiv.org/abs/1704.08699} {arXiv:1704.08699 [hep-th]}
  \BibitemShut {NoStop}%
\bibitem [{\citenamefont {Spaliński}(2018)}]{Spalinski:2017mel}%
  \BibitemOpen
  \bibfield  {author} {\bibinfo {author} {\bibfnamefont {M.}~\bibnamefont
  {Spaliński}},\ }\href {\doibase 10.1016/j.physletb.2017.11.059} {\bibfield
  {journal} {\bibinfo  {journal} {Phys. Lett.}\ }\textbf {\bibinfo {volume}
  {B776}},\ \bibinfo {pages} {468} (\bibinfo {year} {2018})},\ \Eprint
  {http://arxiv.org/abs/1708.01921} {arXiv:1708.01921 [hep-th]} \BibitemShut
  {NoStop}%
\bibitem [{\citenamefont
  {Romatschke}(2017{\natexlab{b}})}]{Romatschke:2017acs}%
  \BibitemOpen
  \bibfield  {author} {\bibinfo {author} {\bibfnamefont {P.}~\bibnamefont
  {Romatschke}},\ }\href {\doibase 10.1007/JHEP12(2017)079} {\bibfield
  {journal} {\bibinfo  {journal} {JHEP}\ }\textbf {\bibinfo {volume} {12}},\
  \bibinfo {pages} {079} (\bibinfo {year} {2017}{\natexlab{b}})},\ \Eprint
  {http://arxiv.org/abs/1710.03234} {arXiv:1710.03234 [hep-th]} \BibitemShut
  {NoStop}%
\bibitem [{\citenamefont {Strickland}\ \emph {et~al.}(2017)\citenamefont
  {Strickland}, \citenamefont {Noronha},\ and\ \citenamefont
  {Denicol}}]{Strickland:2017kux}%
  \BibitemOpen
  \bibfield  {author} {\bibinfo {author} {\bibfnamefont {M.}~\bibnamefont
  {Strickland}}, \bibinfo {author} {\bibfnamefont {J.}~\bibnamefont {Noronha}},
  \ and\ \bibinfo {author} {\bibfnamefont {G.}~\bibnamefont {Denicol}},\
  }\href@noop {} {\  (\bibinfo {year} {2017})},\ \Eprint
  {http://arxiv.org/abs/1709.06644} {arXiv:1709.06644 [nucl-th]} \BibitemShut
  {NoStop}%
\bibitem [{\citenamefont {Florkowski}\ \emph {et~al.}(2017)\citenamefont
  {Florkowski}, \citenamefont {Maksymiuk},\ and\ \citenamefont
  {Ryblewski}}]{Florkowski:2017jnz}%
  \BibitemOpen
  \bibfield  {author} {\bibinfo {author} {\bibfnamefont {W.}~\bibnamefont
  {Florkowski}}, \bibinfo {author} {\bibfnamefont {E.}~\bibnamefont
  {Maksymiuk}}, \ and\ \bibinfo {author} {\bibfnamefont {R.}~\bibnamefont
  {Ryblewski}},\ }\href@noop {} {\  (\bibinfo {year} {2017})},\ \Eprint
  {http://arxiv.org/abs/1710.07095} {arXiv:1710.07095 [hep-ph]} \BibitemShut
  {NoStop}%
\bibitem [{\citenamefont {Denicol}\ and\ \citenamefont
  {Noronha}(2017)}]{Denicol:2017lxn}%
  \BibitemOpen
  \bibfield  {author} {\bibinfo {author} {\bibfnamefont {G.~S.}\ \bibnamefont
  {Denicol}}\ and\ \bibinfo {author} {\bibfnamefont {J.}~\bibnamefont
  {Noronha}},\ }\href@noop {} {\  (\bibinfo {year} {2017})},\ \Eprint
  {http://arxiv.org/abs/1711.01657} {arXiv:1711.01657 [nucl-th]} \BibitemShut
  {NoStop}%
\bibitem [{\citenamefont {Casalderrey-Solana}\ \emph
  {et~al.}(2017)\citenamefont {Casalderrey-Solana}, \citenamefont {Gushterov},\
  and\ \citenamefont {Meiring}}]{Casalderrey-Solana:2017zyh}%
  \BibitemOpen
  \bibfield  {author} {\bibinfo {author} {\bibfnamefont {J.}~\bibnamefont
  {Casalderrey-Solana}}, \bibinfo {author} {\bibfnamefont {N.~I.}\ \bibnamefont
  {Gushterov}}, \ and\ \bibinfo {author} {\bibfnamefont {B.}~\bibnamefont
  {Meiring}},\ }\href@noop {} {\  (\bibinfo {year} {2017})},\ \Eprint
  {http://arxiv.org/abs/1712.02772} {arXiv:1712.02772 [hep-th]} \BibitemShut
  {NoStop}%
\bibitem [{\citenamefont {Blaizot}\ and\ \citenamefont
  {Yan}(2017)}]{Blaizot:2017ucy}%
  \BibitemOpen
  \bibfield  {author} {\bibinfo {author} {\bibfnamefont {J.-P.}\ \bibnamefont
  {Blaizot}}\ and\ \bibinfo {author} {\bibfnamefont {L.}~\bibnamefont {Yan}},\
  }\href@noop {} {\  (\bibinfo {year} {2017})},\ \Eprint
  {http://arxiv.org/abs/1712.03856} {arXiv:1712.03856 [nucl-th]} \BibitemShut
  {NoStop}%
\bibitem [{\citenamefont {Brewer}\ and\ \citenamefont
  {Romatschke}(2015)}]{Brewer:2015ipa}%
  \BibitemOpen
  \bibfield  {author} {\bibinfo {author} {\bibfnamefont {J.}~\bibnamefont
  {Brewer}}\ and\ \bibinfo {author} {\bibfnamefont {P.}~\bibnamefont
  {Romatschke}},\ }\href {\doibase 10.1103/PhysRevLett.115.190404} {\bibfield
  {journal} {\bibinfo  {journal} {Phys. Rev. Lett.}\ }\textbf {\bibinfo
  {volume} {115}},\ \bibinfo {pages} {190404} (\bibinfo {year} {2015})},\
  \Eprint {http://arxiv.org/abs/1508.01199} {arXiv:1508.01199 [hep-th]}
  \BibitemShut {NoStop}%
\bibitem [{\citenamefont {Bantilan}\ \emph {et~al.}(2016)\citenamefont
  {Bantilan}, \citenamefont {Brewer}, \citenamefont {Ishii}, \citenamefont
  {Lewis},\ and\ \citenamefont {Romatschke}}]{Bantilan:2016qos}%
  \BibitemOpen
  \bibfield  {author} {\bibinfo {author} {\bibfnamefont {H.}~\bibnamefont
  {Bantilan}}, \bibinfo {author} {\bibfnamefont {J.~T.}\ \bibnamefont
  {Brewer}}, \bibinfo {author} {\bibfnamefont {T.}~\bibnamefont {Ishii}},
  \bibinfo {author} {\bibfnamefont {W.~E.}\ \bibnamefont {Lewis}}, \ and\
  \bibinfo {author} {\bibfnamefont {P.}~\bibnamefont {Romatschke}},\ }\href
  {\doibase 10.1103/PhysRevA.94.033621} {\bibfield  {journal} {\bibinfo
  {journal} {Phys. Rev.}\ }\textbf {\bibinfo {volume} {A94}},\ \bibinfo {pages}
  {033621} (\bibinfo {year} {2016})},\ \Eprint
  {http://arxiv.org/abs/1605.00014} {arXiv:1605.00014 [cond-mat.quant-gas]}
  \BibitemShut {NoStop}%
\bibitem [{\citenamefont {Grozdanov}\ \emph {et~al.}(2016)\citenamefont
  {Grozdanov}, \citenamefont {Kaplis},\ and\ \citenamefont
  {Starinets}}]{Grozdanov:2016vgg}%
  \BibitemOpen
  \bibfield  {author} {\bibinfo {author} {\bibfnamefont {S.}~\bibnamefont
  {Grozdanov}}, \bibinfo {author} {\bibfnamefont {N.}~\bibnamefont {Kaplis}}, \
  and\ \bibinfo {author} {\bibfnamefont {A.~O.}\ \bibnamefont {Starinets}},\
  }\href {\doibase 10.1007/JHEP07(2016)151} {\bibfield  {journal} {\bibinfo
  {journal} {JHEP}\ }\textbf {\bibinfo {volume} {07}},\ \bibinfo {pages} {151}
  (\bibinfo {year} {2016})},\ \Eprint {http://arxiv.org/abs/1605.02173}
  {arXiv:1605.02173 [hep-th]} \BibitemShut {NoStop}%
\bibitem [{\citenamefont {Bazow}\ \emph {et~al.}(2016)\citenamefont {Bazow},
  \citenamefont {Martinez},\ and\ \citenamefont {Heinz}}]{Bazow:2015zca}%
  \BibitemOpen
  \bibfield  {author} {\bibinfo {author} {\bibfnamefont {D.}~\bibnamefont
  {Bazow}}, \bibinfo {author} {\bibfnamefont {M.}~\bibnamefont {Martinez}}, \
  and\ \bibinfo {author} {\bibfnamefont {U.~W.}\ \bibnamefont {Heinz}},\ }\href
  {\doibase 10.1103/PhysRevD.93.034002} {\bibfield  {journal} {\bibinfo
  {journal} {Phys. Rev.}\ }\textbf {\bibinfo {volume} {D93}},\ \bibinfo {pages}
  {034002} (\bibinfo {year} {2016})},\ \bibinfo {note} {[Phys.
  Rev.D93,034002(2016)]},\ \Eprint {http://arxiv.org/abs/1507.06595}
  {arXiv:1507.06595 [nucl-th]} \BibitemShut {NoStop}%
\bibitem [{Note1()}]{Note1}%
  \BibitemOpen
  \bibinfo {note} {M.~Martinez and U.~Heinz, private
  communication.}\BibitemShut {Stop}%
\bibitem [{\citenamefont {Dunne}()}]{Dunne}%
  \BibitemOpen
  \bibfield  {author} {\bibinfo {author} {\bibfnamefont {G.~V.}\ \bibnamefont
  {Dunne}},\ }\href@noop {} {\bibinfo  {journal} {Lectures given at the
  Schladming Winter School 2015}\ }\BibitemShut {NoStop}%
\bibitem [{\citenamefont {Dunne}\ and\ \citenamefont
  {Ünsal}(2016)}]{Dunne:2015eaa}%
  \BibitemOpen
\bibfield  {journal} {  }\bibfield  {author} {\bibinfo {author} {\bibfnamefont
  {G.~V.}\ \bibnamefont {Dunne}}\ and\ \bibinfo {author} {\bibfnamefont
  {M.}~\bibnamefont {Ünsal}},\ }\bibfield  {booktitle} {\emph {\bibinfo
  {booktitle} {{Proceedings, 33rd International Symposium on Lattice Field
  Theory (Lattice 2015): Kobe, Japan, July 14-18, 2015}}},\ }\href@noop {}
  {\bibfield  {journal} {\bibinfo  {journal} {PoS}\ }\textbf {\bibinfo {volume}
  {LATTICE2015}},\ \bibinfo {pages} {010} (\bibinfo {year} {2016})},\ \Eprint
  {http://arxiv.org/abs/1511.05977} {arXiv:1511.05977 [hep-lat]} \BibitemShut
  {NoStop}%
\bibitem [{\citenamefont {Basar}\ \emph {et~al.}(2013)\citenamefont {Basar},
  \citenamefont {Dunne},\ and\ \citenamefont {Unsal}}]{Basar:2013eka}%
  \BibitemOpen
  \bibfield  {author} {\bibinfo {author} {\bibfnamefont {G.}~\bibnamefont
  {Basar}}, \bibinfo {author} {\bibfnamefont {G.~V.}\ \bibnamefont {Dunne}}, \
  and\ \bibinfo {author} {\bibfnamefont {M.}~\bibnamefont {Unsal}},\ }\href
  {\doibase 10.1007/JHEP10(2013)041} {\bibfield  {journal} {\bibinfo  {journal}
  {JHEP}\ }\textbf {\bibinfo {volume} {10}},\ \bibinfo {pages} {041} (\bibinfo
  {year} {2013})},\ \Eprint {http://arxiv.org/abs/1308.1108} {arXiv:1308.1108
  [hep-th]} \BibitemShut {NoStop}%
\bibitem [{\citenamefont {Aniceto}\ \emph {et~al.}(2018)\citenamefont
  {Aniceto}, \citenamefont {Ba¸sarba¸sar},\ and\ \citenamefont
  {Schiappa}}]{Aniceto:2018bis}%
  \BibitemOpen
  \bibfield  {author} {\bibinfo {author} {\bibfnamefont {I.}~\bibnamefont
  {Aniceto}}, \bibinfo {author} {\bibfnamefont {G.}~\bibnamefont
  {Ba¸sarba¸sar}}, \ and\ \bibinfo {author} {\bibfnamefont {R.}~\bibnamefont
  {Schiappa}},\ }\href@noop {} {\  (\bibinfo {year} {2018})},\ \Eprint
  {http://arxiv.org/abs/1802.10441} {arXiv:1802.10441 [hep-th]} \BibitemShut
  {NoStop}%
\bibitem [{\citenamefont {Jankowski}\ \emph {et~al.}(2014)\citenamefont
  {Jankowski}, \citenamefont {Plewa},\ and\ \citenamefont
  {Spalinski}}]{Jankowski:2014lna}%
  \BibitemOpen
  \bibfield  {author} {\bibinfo {author} {\bibfnamefont {J.}~\bibnamefont
  {Jankowski}}, \bibinfo {author} {\bibfnamefont {G.}~\bibnamefont {Plewa}}, \
  and\ \bibinfo {author} {\bibfnamefont {M.}~\bibnamefont {Spalinski}},\ }\href
  {\doibase 10.1007/JHEP12(2014)105} {\bibfield  {journal} {\bibinfo  {journal}
  {JHEP}\ }\textbf {\bibinfo {volume} {12}},\ \bibinfo {pages} {105} (\bibinfo
  {year} {2014})},\ \Eprint {http://arxiv.org/abs/1411.1969} {arXiv:1411.1969
  [hep-th]} \BibitemShut {NoStop}%
\bibitem [{\citenamefont {Keegan}\ \emph {et~al.}(2015)\citenamefont {Keegan},
  \citenamefont {Kurkela}, \citenamefont {Romatschke}, \citenamefont {van~der
  Schee},\ and\ \citenamefont {Zhu}}]{Keegan:2015avk}%
  \BibitemOpen
  \bibfield  {author} {\bibinfo {author} {\bibfnamefont {L.}~\bibnamefont
  {Keegan}}, \bibinfo {author} {\bibfnamefont {A.}~\bibnamefont {Kurkela}},
  \bibinfo {author} {\bibfnamefont {P.}~\bibnamefont {Romatschke}}, \bibinfo
  {author} {\bibfnamefont {W.}~\bibnamefont {van~der Schee}}, \ and\ \bibinfo
  {author} {\bibfnamefont {Y.}~\bibnamefont {Zhu}},\ }\href@noop {} {\
  (\bibinfo {year} {2015})},\ \Eprint {http://arxiv.org/abs/1512.05347}
  {arXiv:1512.05347 [hep-th]} \BibitemShut {NoStop}%
\bibitem [{\citenamefont {Bjorken}(1983)}]{Bjorken:1982qr}%
  \BibitemOpen
  \bibfield  {author} {\bibinfo {author} {\bibfnamefont {J.}~\bibnamefont
  {Bjorken}},\ }\href {\doibase 10.1103/PhysRevD.27.140} {\bibfield  {journal}
  {\bibinfo  {journal} {Phys.Rev.}\ }\textbf {\bibinfo {volume} {D27}},\
  \bibinfo {pages} {140} (\bibinfo {year} {1983})}\BibitemShut {NoStop}%
\bibitem [{\citenamefont {Baym}(1984)}]{Baym:1984np}%
  \BibitemOpen
  \bibfield  {author} {\bibinfo {author} {\bibfnamefont {G.}~\bibnamefont
  {Baym}},\ }\href {\doibase 10.1016/0370-2693(84)91863-X} {\bibfield
  {journal} {\bibinfo  {journal} {Phys. Lett.}\ }\textbf {\bibinfo {volume}
  {B138}},\ \bibinfo {pages} {18} (\bibinfo {year} {1984})}\BibitemShut
  {NoStop}%
\bibitem [{\citenamefont {Florkowski}\ \emph {et~al.}(2013)\citenamefont
  {Florkowski}, \citenamefont {Ryblewski},\ and\ \citenamefont
  {Strickland}}]{Florkowski:2013lya}%
  \BibitemOpen
  \bibfield  {author} {\bibinfo {author} {\bibfnamefont {W.}~\bibnamefont
  {Florkowski}}, \bibinfo {author} {\bibfnamefont {R.}~\bibnamefont
  {Ryblewski}}, \ and\ \bibinfo {author} {\bibfnamefont {M.}~\bibnamefont
  {Strickland}},\ }\href {\doibase 10.1103/PhysRevC.88.024903} {\bibfield
  {journal} {\bibinfo  {journal} {Phys. Rev.}\ }\textbf {\bibinfo {volume}
  {C88}},\ \bibinfo {pages} {024903} (\bibinfo {year} {2013})},\ \Eprint
  {http://arxiv.org/abs/1305.7234} {arXiv:1305.7234 [nucl-th]} \BibitemShut
  {NoStop}%
\bibitem [{\citenamefont {Jaiswal}(2013)}]{Jaiswal:2013npa}%
  \BibitemOpen
  \bibfield  {author} {\bibinfo {author} {\bibfnamefont {A.}~\bibnamefont
  {Jaiswal}},\ }\href {\doibase 10.1103/PhysRevC.87.051901} {\bibfield
  {journal} {\bibinfo  {journal} {Phys. Rev.}\ }\textbf {\bibinfo {volume}
  {C87}},\ \bibinfo {pages} {051901} (\bibinfo {year} {2013})},\ \Eprint
  {http://arxiv.org/abs/1302.6311} {arXiv:1302.6311 [nucl-th]} \BibitemShut
  {NoStop}%
\bibitem [{\citenamefont {Buchel}\ \emph {et~al.}(2016)\citenamefont {Buchel},
  \citenamefont {Heller},\ and\ \citenamefont {Noronha}}]{Buchel:2016cbj}%
  \BibitemOpen
  \bibfield  {author} {\bibinfo {author} {\bibfnamefont {A.}~\bibnamefont
  {Buchel}}, \bibinfo {author} {\bibfnamefont {M.~P.}\ \bibnamefont {Heller}},
  \ and\ \bibinfo {author} {\bibfnamefont {J.}~\bibnamefont {Noronha}},\
  }\href@noop {} {\  (\bibinfo {year} {2016})},\ \Eprint
  {http://arxiv.org/abs/1603.05344} {arXiv:1603.05344 [hep-th]} \BibitemShut
  {NoStop}%
\bibitem [{\citenamefont {Basar}\ and\ \citenamefont
  {Dunne}(2015)}]{Basar:2015ava}%
  \BibitemOpen
  \bibfield  {author} {\bibinfo {author} {\bibfnamefont {G.}~\bibnamefont
  {Basar}}\ and\ \bibinfo {author} {\bibfnamefont {G.~V.}\ \bibnamefont
  {Dunne}},\ }\href {\doibase 10.1103/PhysRevD.92.125011} {\bibfield  {journal}
  {\bibinfo  {journal} {Phys. Rev.}\ }\textbf {\bibinfo {volume} {D92}},\
  \bibinfo {pages} {125011} (\bibinfo {year} {2015})},\ \Eprint
  {http://arxiv.org/abs/1509.05046} {arXiv:1509.05046 [hep-th]} \BibitemShut
  {NoStop}%
\bibitem [{\citenamefont {Aniceto}\ and\ \citenamefont
  {Spaliński}(2016)}]{Aniceto:2015mto}%
  \BibitemOpen
  \bibfield  {author} {\bibinfo {author} {\bibfnamefont {I.}~\bibnamefont
  {Aniceto}}\ and\ \bibinfo {author} {\bibfnamefont {M.}~\bibnamefont
  {Spaliński}},\ }\href {\doibase 10.1103/PhysRevD.93.085008} {\bibfield
  {journal} {\bibinfo  {journal} {Phys. Rev.}\ }\textbf {\bibinfo {volume}
  {D93}},\ \bibinfo {pages} {085008} (\bibinfo {year} {2016})},\ \Eprint
  {http://arxiv.org/abs/1511.06358} {arXiv:1511.06358 [hep-th]} \BibitemShut
  {NoStop}%
\bibitem [{\citenamefont {{Yamada}}\ and\ \citenamefont
  {{Ikeda}}(2013)}]{PadeCut}%
  \BibitemOpen
  \bibfield  {author} {\bibinfo {author} {\bibfnamefont {H.~S.}\ \bibnamefont
  {{Yamada}}}\ and\ \bibinfo {author} {\bibfnamefont {K.~S.}\ \bibnamefont
  {{Ikeda}}},\ }\href@noop {} {\  (\bibinfo {year} {2013})},\ \Eprint
  {http://arxiv.org/abs/1308.4453} {arXiv:1308.4453 [math-ph]} \BibitemShut
  {NoStop}%
\bibitem [{\citenamefont {Florkowski}\ \emph {et~al.}(2016)\citenamefont
  {Florkowski}, \citenamefont {Ryblewski},\ and\ \citenamefont
  {Spalinski}}]{Florkowski:2016zsi}%
  \BibitemOpen
  \bibfield  {author} {\bibinfo {author} {\bibfnamefont {W.}~\bibnamefont
  {Florkowski}}, \bibinfo {author} {\bibfnamefont {R.}~\bibnamefont
  {Ryblewski}}, \ and\ \bibinfo {author} {\bibfnamefont {M.}~\bibnamefont
  {Spalinski}},\ }\href@noop {} {\  (\bibinfo {year} {2016})},\ \Eprint
  {http://arxiv.org/abs/1608.07558} {arXiv:1608.07558 [nucl-th]} \BibitemShut
  {NoStop}%
\bibitem [{\citenamefont {Janik}\ and\ \citenamefont
  {Peschanski}(2006)}]{Janik:2006gp}%
  \BibitemOpen
  \bibfield  {author} {\bibinfo {author} {\bibfnamefont {R.~A.}\ \bibnamefont
  {Janik}}\ and\ \bibinfo {author} {\bibfnamefont {R.~B.}\ \bibnamefont
  {Peschanski}},\ }\href {\doibase 10.1103/PhysRevD.74.046007} {\bibfield
  {journal} {\bibinfo  {journal} {Phys. Rev.}\ }\textbf {\bibinfo {volume}
  {D74}},\ \bibinfo {pages} {046007} (\bibinfo {year} {2006})},\ \Eprint
  {http://arxiv.org/abs/hep-th/0606149} {arXiv:hep-th/0606149 [hep-th]}
  \BibitemShut {NoStop}%
\bibitem [{\citenamefont {Heller}\ \emph {et~al.}(2014)\citenamefont {Heller},
  \citenamefont {Janik}, \citenamefont {Spaliński},\ and\ \citenamefont
  {Witaszczyk}}]{Heller:2014wfa}%
  \BibitemOpen
  \bibfield  {author} {\bibinfo {author} {\bibfnamefont {M.~P.}\ \bibnamefont
  {Heller}}, \bibinfo {author} {\bibfnamefont {R.~A.}\ \bibnamefont {Janik}},
  \bibinfo {author} {\bibfnamefont {M.}~\bibnamefont {Spaliński}}, \ and\
  \bibinfo {author} {\bibfnamefont {P.}~\bibnamefont {Witaszczyk}},\ }\href
  {\doibase 10.1103/PhysRevLett.113.261601} {\bibfield  {journal} {\bibinfo
  {journal} {Phys.Rev.Lett.}\ }\textbf {\bibinfo {volume} {113}},\ \bibinfo
  {pages} {261601} (\bibinfo {year} {2014})},\ \Eprint
  {http://arxiv.org/abs/1409.5087} {arXiv:1409.5087 [hep-th]} \BibitemShut
  {NoStop}%
\bibitem [{\citenamefont {Heller}\ and\ \citenamefont
  {Svensson}(2018)}]{Heller:2018qvh}%
  \BibitemOpen
  \bibfield  {author} {\bibinfo {author} {\bibfnamefont {M.~P.}\ \bibnamefont
  {Heller}}\ and\ \bibinfo {author} {\bibfnamefont {V.}~\bibnamefont
  {Svensson}},\ }\href@noop {} {\  (\bibinfo {year} {2018})},\ \Eprint
  {http://arxiv.org/abs/1802.08225} {arXiv:1802.08225 [nucl-th]} \BibitemShut
  {NoStop}%
\bibitem [{\citenamefont {Aniceto}(2016)}]{Aniceto:2015rua}%
  \BibitemOpen
  \bibfield  {author} {\bibinfo {author} {\bibfnamefont {I.}~\bibnamefont
  {Aniceto}},\ }\href {\doibase 10.1088/1751-8113/49/6/065403} {\bibfield
  {journal} {\bibinfo  {journal} {J. Phys.}\ }\textbf {\bibinfo {volume}
  {A49}},\ \bibinfo {pages} {065403} (\bibinfo {year} {2016})},\ \Eprint
  {http://arxiv.org/abs/1506.03388} {arXiv:1506.03388 [hep-th]} \BibitemShut
  {NoStop}%
\bibitem [{\citenamefont {Heller}\ \emph
  {et~al.}(2012{\natexlab{b}})\citenamefont {Heller}, \citenamefont {Janik},\
  and\ \citenamefont {Witaszczyk}}]{Heller:2012je}%
  \BibitemOpen
  \bibfield  {author} {\bibinfo {author} {\bibfnamefont {M.~P.}\ \bibnamefont
  {Heller}}, \bibinfo {author} {\bibfnamefont {R.~A.}\ \bibnamefont {Janik}}, \
  and\ \bibinfo {author} {\bibfnamefont {P.}~\bibnamefont {Witaszczyk}},\
  }\href {\doibase 10.1103/PhysRevD.85.126002} {\bibfield  {journal} {\bibinfo
  {journal} {Phys. Rev.}\ }\textbf {\bibinfo {volume} {D85}},\ \bibinfo {pages}
  {126002} (\bibinfo {year} {2012}{\natexlab{b}})},\ \Eprint
  {http://arxiv.org/abs/1203.0755} {arXiv:1203.0755 [hep-th]} \BibitemShut
  {NoStop}%
\end{thebibliography}%
\end{document}